\documentclass[aps,nofootinbib,showpacs,twocolumn,longbibliography]{revtex4-1}
\usepackage{amstext,amsmath,amssymb,amsfonts,bbm}
\usepackage[latin1]{inputenc}
\usepackage{graphicx}
\usepackage{epstopdf}
\usepackage{amsthm}
\usepackage{tocvsec2}
\usepackage{enumerate}
\usepackage{subfigure}
\usepackage{color}

\usepackage{multirow} \usepackage{rotating}
\usepackage[colorlinks,citecolor=blue,linkcolor=blue,urlcolor=blue]{hyperref}

\def\beq{\begin{equation}}
\def\be{\begin{equation}}
\def\ee{\end{equation}}
\def\bes{\begin{eqnarray}}
\def\ees{\end{eqnarray}}

\def\R{{\mathbbm R}}

\begin{document}
\maxtocdepth{subsection}
%%%%%%%%%%%%%%%%%%%%%%%%%%%%%%%%%%%%%%%%%%%%%%%%%%%

\title{\large \bf On the structure of a background independent quantum theory:\\ 
Hamilton function, transition amplitudes, classical limit and continuous limit}
\author{{\bf Carlo Rovelli}}
\affiliation{Centre de Physique Th\'eorique, Case 907,  Luminy, F-13288 Marseille, EU}

\date{\small\today}

%%%%%%%%%%%%%%%%%%%%%%%%%%%%%%%%%%%%%
\begin{abstract}\noindent
The Hamilton function is a powerful tool for studying the classical limit of quantum systems, which remains 
meaningful in background-independent systems.  In quantum gravity, it clarifies the physical interpretation of the transitions amplitudes and their truncations.

\end{abstract} 
%%%%%%%%%%%%%%%%%%%%%%%%%%%%%%%%%%%%%%

%\pacs{04.60.Pp}
\maketitle

%%%%%%%%%%%%%%%%%%%%%%%%%%%%%%%%%%%%%%

\section{Systems evolving in time} 

Consider a dynamical system with configuration variable  $q\!\in\!{\cal C}$, and lagrangian $L(q,\dot q)$. Given an initial configuration $q$ at time $t$ and a final configuration $q'$ at time $t'$, let $q_{q,t,q',t'}:{\R}\to {\cal C}$ be a solution of the equations of motion such $q_{q,t,q',t'}(t)=q$ and $q_{q,t,q',t'}(t')=q'$. Assume for the moment this exists and is unique.  The Hamilton function is the function on  $({\cal C}\times \R)^2$ defined by 
\be
   S({q,t,q',t'})=\int_t^{t'} dt \ L(q_{q,t,q',t'},\dot q_{q,t,q',t'}),      \label{one}
\ee
namely the value of the action on the solution of the equation of motion determined by given initial and final data.  This function, introduced by Hamilton in 1834 \cite{Hamilton:1834uq} codes the solution of the dynamics of the system, has  remarkable properties and is a powerful tool that remains meaningful in background-independent physics. 

Let $H$ be the quantum hamiltonian operator of the system and $|q\rangle$ the eigenstates of its $q$ observables. The transition amplitude 
\be
  W({q,t,q',t'})=\langle q'|e^{-\frac{i}{\hbar}H(t'-t)}|q\rangle.
\ee
codes all the quantum dynamics. In a path integral formulation, it can be written as 
\be
  W({q,t,q',t'})=\int_{q(t)=q}^{q(t')=q'} D[q]\ e^{\frac{i}{\hbar}\int_t^{t'} dt L(q,\dot q)}. \label{fi}
\ee
In the limit in which $\hbar$ can be considered small, this can be evaluated by a saddle point approximation, and gives
\be
  W({q,t,q',t'})\sim e^{\frac{i}{\hbar}S({q,t,q',t'})}.
\ee
That is, the classical limit of the quantum theory can be obtained by reading out the Hamilton function from the quantum transition amplitude:
\be
  \lim_{\hbar\to0} (-i\hbar) \log W({q,t,q',t'})=S({q,t,q',t'}).
\ee 
The functional integral in \eqref{fi} can be  defined either by perturbation theory around a gaussian integral, or as a limit of multiple integrals. Let us focus on the second definition, useful in non-perturbative theories such as lattice QCD and quantum gravity, which are not defined by a gaussian point.  Let $L(q_n,q_{n-1},t_n,t_{n-1})$ be a discretization of the lagrangian. The multiple integral 
\be
  W_N({q,t,q',t'})=\int \frac{dq_n}{\mu(q_n)} \,  e^{\frac{i}{\hbar}\sum_{n=1}^{N} a L(q_n,q_{n\!-\!1},t_n,t_{n\!-\!1})} \label{di} 
\ee
where $\mu(q_n)$ is a suitable measure factor, 
$
t_n\!=\!n(t'\!-\!t)/{N} \equiv n\,a,
$ and the boundary data are $q_0=q$ and $q_N=q'$, has two distinct limits. The \emph{continuous limit}
\be
  \lim_{N\to\infty}W_N({q,t,q',t'})=W({q,t,q',t'})
\ee
gives the transition amplitude. While the \emph{classical limit}
\be
  \lim_{\hbar\to0} (-i\hbar) \log W_N({q,t,q',t'})=S_N({q,t,q',t'}).
\ee 
gives the Hamilton function of the classical discretized  system, namely the value of the action $\sum_{n=1}^{N} aL(q_n,q_{n-1},t_n,t_{n-1})$ on the sequence $q_n$ that extremizes this action at given boundary data.  The discretization is good if the classical theory is recovered as the continuous limit of the discretized theory, that is, if
\be
  \lim_{N\to\infty}S_N({q,t,q',t'})=S({q,t,q',t'}).
\ee
Summarizing:

\begin{table}[h]
  \centering 
\multirow{2}{10mm}{\begin{sideways}\parbox{20mm}{$\xrightarrow{ \hspace*{2em}{\rm Continuous\ limit} \hspace*{1em}}$}\end{sideways}}
  \begin{tabular}{@{} ccc @{}}
   \parbox{2.8cm}{\scriptsize  Quantum theory \\  \emph{Transition amplitude}\\ $W({q,t,q',t'})$} \hspace*{1em}& $\xrightarrow{\hbar\to0}$ &\hspace*{1em} \parbox{2.5cm}{\scriptsize Classical theory\\ \emph{Hamilton function} \\ $S({q,t,q',t'})$} \\[4mm]
{\begin{sideways}\parbox{10mm}{ $\xrightarrow{{ N}\to\infty}$}\end{sideways}} \hspace*{1em}&  &\hspace*{1em} {\begin{sideways}\parbox{10mm}{ $\xrightarrow{{ N}\to\infty}$}\end{sideways}}  \\[1mm]
   \parbox{2cm}{ \scriptsize Discretized \\ quantum theory $W_N({q,t,q',t'})$}\hspace*{1em} & $\xrightarrow{\hbar\to0}$ & \hspace*{1em}\parbox{2cm}{\scriptsize Discretized classical theory\\  $S_N({q,t,q',t'})$} \\ 
  \end{tabular}\\[.2cm]
\parbox{35mm}{$\xrightarrow{ \hspace*{4em}{\rm Classical\ limit} \hspace*{4em}}$}
  \label{tab:label}
    \caption{Continuous and classical limits}
\end{table}

The interest of this structure is that it remains meaningful in diffeomorphism invariant systems and offers an excellent conceptual tool for dealing with background independent physics. To see this, let's first consider its generalization to finite dimensional parametrized systems. 

\section{Parametrized systems} 

I start by revising a few well-known facts about background independence. The system considered above can be equivalently reformulated as follows. Introduce a new evolution parameter $\tau$ and promote the physical time $t$ to a new dynamical variable $t(\tau)$.  By changing variables $q(t)\to q(\tau)=q(t(\tau))$ in the action, we obtain a new action which depends on $n+1$ variables $x=(q,t)\in{\cal C\times \R}\equiv{\cal C}_{ex}$ evolving in $\tau$
\be
I[q]\!=\!\!\int \!dt\, L(q,\dot q) \to \!
\int\! d\tau\,  \dot t\ L(q,\dot q/\dot t)=\!\!\int\! d\tau\,\! {\cal L}(x,\dot x)  \equiv I[x]
\ee
where on the l.h.s of the arrow the dot indicates a derivative with respect to $t$, while on the r.h.s  of the arrow (and from now on) it indicates a derivative with respect to $\tau$.  

For instance, a single particle lagrangian $L=m\dot q^2/2-V(q)$ gives ${\cal L}=m\dot q^2/(2\dot t)-\dot t V(q)$. This  ``parametrized"  system describes the same physics as the original one,  but in a  different logic. The physical time variable $t$ is now treated on the same footing as the other dynamical variables\footnote{That is, dynamics is not anymore interpreted as the description of the evolution  in $t$ of the $n$ configuration variables $q$, but rather as a description of the possible \emph{relations} between the $n+1$ variables $x=(q,t)$ (see Section 3.2.4 in \cite{Rovelli:2004fk} and \cite{Rovelli:2009ee,Rovelli:2001bz}).} and the system has a local gauge invariance under reparametrizations of $\tau$.  It is easy to see that the canonical Hamiltonian of the action $I[x]$ vanishes and the dynamics is entirely given by the first class constraint
\be
    C(q,t,p,p_t)=p_t+H(q,p)\sim 0.  \label{ham}
\ee
where $p$ and $p_t$ are the momenta conjugate to $q$ and $t$ and $H(q,p)$ is the Hamiltonian of the original system.

As was recognized in the early days of the canonical analysis of general relativity \cite{Deser:1961zza}, this is precisely the structure of general relativity (GR). The general relativistic coordinates $x^\mu$ play the same role as $\tau$ above and reparametrization invariance is the invariance under general coordinate transformations.   Above, the ``parametrized" form of the dynamics has been derived from an original ``un-parametrized" form, while GR is directly written in the parametrized form, with a local gauge-invariance.   ``De-parametrizing" GR is possible in principle, but spoils many of its formal properties, making the formalism far more cumbersome and intractable. In GR we better live with reparametrization invariance and, in fact, take advantage of it. 

What is the Hamilton function of the parametrized system? From the definition
\be
 S(x,\tau,x',\tau')=\int_{\tau}^{\tau'} d\tau\ {\cal L}(x_{x,x'},\dot x_{x,x'}) .
\ee
where $x_{x,x'}(\tau)$ is a solution of the equations of motion such that $x_{x,x'}(\tau)=x$ and $x_{x,x'}(\tau')=x'$.
But a moment of reflection will convince the reader of two facts which are at first surprising:  first, 
\be
S(x,\tau,x',\tau')=S(x,x'), 
\ee
namely the Hamilton function of a parametrized system is independent from $\tau$ and $\tau'$.\footnote{$S$ is a solution of the Hamilton-Jacobi equation 
\be
{\partial S}/{\partial\tau}=H\!\left(\! x,{\partial S}/{\partial x}\right).
\ee
But the canonical Hamiltonian $H(x,p_x)$ vanishes because of the gauge invariance and therefore ${\partial S}/{\partial\tau}=0$.} Second, because of the gauge there are many solutions of the classical equations with the same boundary data, but $S(x,x')$ is independent from the one chosen, that is, is gauge invariant.  Both facts are immediate consequence of the invariance of the action under reparametrizations of $\tau$.  Furthermore, since the new action is just the old one in new variables,  $S(x,x')$ is precisely nothing else than \eqref{one}. 

That is, remarkably, the Hamilton function of the parametrized system is the same object as the Hamilton function of the original system. In fact, notice that the original Hamilton function was already a function of (two copies of) the \emph{extended} configuration space ${\cal C}_{ex}={\cal C\times \R}$.  The Hamilton function was already born as if it knew it had to work in a parametrized language!

But this prescience goes, in facts, much farther.  The essential property of the Hamilton function  \eqref{one} is that its derivative gives the momentum\footnote{This is why the Hamilton function contains the solution of the equations of motion. Inverting $p=p(q,t,q',t')$ we have the final position as a function of the initial position and momenta, and of the lapsed time: $q'(t')=q'(q,p,t,t')$, namely the solution of the equations of motion.}  
\be
   \frac{\partial S({q,t,q',t'})}{\partial q}=p(q,t,q',t')
\ee
But, a bit magically, it also happens to be true that 
\be
   \frac{\partial S({q,t,q',t'})}{\partial t}=-E(q,t,q',t')
\ee
where $E$ is the energy at time $t$ on the given trajectory, and the energy is precisely $-p_t$, as it is clear from \eqref{ham}. Therefore the Hamilton function of the unparametrized system already treats $q$ and $t$ on the same footing.

Let me now come to the discretization of the path integral.  What happens if we discretize the path integral of the reparametrization invariant theory?  This can be achieved by a short-cut as follows. Instead of fixing the discretization of the time steps by $t_n=na$ in \eqref{di}, let's integrate over the positions of the steps
\be
  W_N(x,x')=\int \frac{dq_n dt_n}{\mu(q_n,t_n)} \,  e^{\frac{i}{\hbar}\sum_{n=1}^{N} (t_n\!-t_{n\!-\!1}) L(q_n,q_{n\!-\!1},t_n,t_{n\!-\!1})} \label{dip}, 
\ee
Notice that the ``lattice spacing" $a$ is replaced by the variable quantity $(t_n-t_{n-1})$. With an appropriate choice of integration range and measure ${\mu(q_n,t_n)}$, this discretization provides a viable alternative to \eqref{di} \cite{Rovelli:2011fk}.  

We can therefore recover the structure illustrated in Table I.  The discretized quantum transition amplitude \eqref{dip} yields the full quantum transition amplitude in the $N\to\infty$ limit and the Hamilton function of a discretization of the classical parametrized theory in the $\hbar\to 0$ limit. A concrete example is illustrated in detail in \cite{Rovelli:2011fk}. 

\section{Three important points} 

Before applying these ideas to quantum gravity, three observations are required.

\emph{(i)} I have assumed in the first paragraph of this article that there is one and only one solution of the equations of motion for any given boundary data.  This is obviously not true in general. Boundary data can bound several solutions, or no solutions at all.   A simple case is when the data can bound  \emph{gauge equivalent} solutions, which have the same action. This has no effect on the definition of the Hamilton function, which is gauge invariant.  Consider then  the cases in which there are gauge-inequivalent solutions, or no solution at all. 

A moment of reflection will convince the reader that the existence of these cases does not modify the structure of Table I. The fundamental object is the quantum transition amplitude.  The Hamilton function is a derived quantity that emerges from the classical limit of the transition amplitude.  If there is more than a single classical trajectory between certain boundary data, then the saddle point approximation of the functional integral will give a sum over different critical points of the action.  If there is no classical solution at all, then the integral will be suppressed in the classical limit. Therefore we simply interpret the Hamilton function as a function on ${\cal C}_{ex}^2$ which on some regions can be multivalued, or non defined. 

\emph{(ii)} Care is required with boundary terms in the action. A total derivative added to the classical action does not  modify the equations of motion but changes the Hamilton function. The form of the action needed to define the Hamilton function depends only on the variables and their first derivatives.  In gravity, the usual action $\int \sqrt{g}R$  vanishes on all pure gravity solutions, but the Hamilton function of GR does not vanish identically. We have to use a proper action that depends only on the metric and its first derivatives, or, equivalently, include the proper boundary term, as we do below.  

\emph{(iii)} In a finite dimensional system as the one considered above, the boundary data are defined on the boundary of a finite portion of a physical trajectory.  The generalization of this procedure useful in field theory and in quantum gravity is to consider a \emph{finite} portion of a physical trajectory as well.  For this, consider a \emph{finite} region $\cal B$ of spacetime and its boundary $\Sigma=\partial{\cal B}$.  For a field theory on a metric space, the analog of the boundary data $(q,q')$ is given by the fields value on $\Sigma$, while the analog of the data $(t,t')$, that determine the location of the boundary in the background time $t$, is given by the coordinate position $\Sigma: \sigma\to x^\mu(\sigma)$ of the boundary 3d surface $\Sigma$ in the background metric spacetime.   

In a general covariant theory, on the other hand, the general relativistic coordinates $x^\mu$ drop out of the Hamilton function, like $\tau$ and $\tau'$ drop from $S(x,\tau,x',\tau')$.  The relevant boundary data are then solely the fields on $\Sigma$.  In particular, in pure gravity, the boundary data can be taken to be just the canonical variable on $\Sigma$. In the ADM formalism the boundary data is given solely by the 3-metric $q$ of $\Sigma$. Notice that in this case the geometrical shape of the boundary surface $\Sigma$ is not determined by its coordinates, but rather by the gravitational field on it, which determines its metric \cite{Modesto:2005sj}.

\section{General relativity} 

Let me now apply the ideas illustrated above to quantum gravity. I start with the Hamilton function of classical GR.  Following the discussion in the previous section, consider a finite 4d region  $\cal B$ surrounded by a 3d surface $\Sigma$.\footnote{In the applications we are particularly interested in the cases where  $\cal B$ is a ball or a segment of a cylinder, and therefore $\Sigma$ has the topology of $S_3$, or $S_3\times S_3$, which are respectively relevant to discuss scattering and cosmology.}  For simplicity, I begin with metric variables -- later on I will use other variables to describe the gravitational field.  We can fix pure gravity boundary data on $\Sigma$ by giving the three metric $q$ of $\Sigma$. Let  $g_q$ be a 4d metric on  $\cal B$ which satisfies the Einstein's equations and induces the 3-metric $q$ on $\Sigma$.  The possibility that  such $g_q$ might be non-existent or not-unique does not concern us here, as discussed in point\,\emph{(i)} of previous section.  Consider the action of GR, including the boundary term 
\be
 I[g]=\frac12 \int_{\cal B} d^4x\ \sqrt{g}R[g]+\int_{\Sigma} d^3x\ \sqrt{q} k,
\ee
where $k=k^{ab}q_{ab}$ is the trace of the extrinsic curvature $k^{ab}$ on the boundary and $q$ is the induced 3 metric.  If $g$ is a solution of the equations of motion, the first term vanishes. Thus, the Hamilton function of $q$ is the action of $g_q$, that is 
\be
 S[q]=I[g_q]=\int_{\Sigma} d^3x\ \sqrt{q}\ k^{ab}[q](x)q_{ab}(x).       \label{HRegee}
\ee
The non-trivial part of this expression is the dependence of the extrinsic curvature $k^{ab}[q](x)$ on the 3-metric.  This dependence is nonlocal: in general the extrinsic curvature in a point $x$ depends on the value of the metric on the entire surface $\Sigma$.\footnote{For a simple example of this dependence consider the Euclidean theory, and say that the metric $q$ is that of a metric 3-sphere with radius $a$.  Such a sphere can be imbedded in many curved 4d manifolds, and  the extrinsic curvature of the imbedding is \emph{not} determined by $q$. But if the 4d  Riemannian manifold is a solution of the Euclidean Einstein equations, then this freedom is drastically reduced. A solution of the Einstein equation is flat space. A metric 3-sphere can be imbedded in a flat 4d space as the surface of the ball with radius $a$ and this fixes the extrinsic curvature to be $k^{ab}=a^{-1}q^{ab}$, so that $H[q(a)]=6\pi^2 a^2$.}  

Now, consider a discretization of GR. As a warm-up, consider the best-known discretization, which is  Regge calculus.  Fix a triangulation $\Delta$ of $\cal B$ and consider Regge geometries on $\cal B$. A Regge geometry is a geometry which is everywhere flat except on the triangles $t$ of $\Delta$, where the curvature is distributional and fully determined by the deficit angle $\theta_t$ at the triangle, determined by the sum of the dihedral angles of the flat 4-simplices meeting at $t$.  This geometry is uniquely determined by giving the lengths $l_i$ of the segments $i$ of $\Delta$, which determine the deficit angles $\theta_t=\theta_t(l_i)$ and the areas $A_t=A_t(l_i)$  of these triangles.  The Regge action is a local function of these lengths and reads
\be
 I_\Delta[l_i]=\sum_{t\in\Delta} \  \theta_t(l_i) A_t(l_i).
\ee
The triangulation of $\cal B$ indices a boundary triangulation $\partial\Delta$ on the boundary surface $\Sigma$. The boundary 3-metric is fully determined by the lengths of the boundary links $i_b$, which I denote as $q=\{l_{i_b}\}$. The Hamilton function of the Regge theory is the discrete analog of \eqref{HRegee}:
\be
 S_{\Delta}[q]=\sum_{t\in\partial\Delta} \  \theta_t[l_{i_b}] A_t(l_{i_b}).
\ee
where the sum is here over the triangles in the boundary and $\theta_t$ is the discrete extrinsic curvature at the boundary triangles, namely the angle between the 4-normals of the two boundary tetrahedra separated by the triangle $t$.  While the area of a boundary triangle depends only on the length of its three sides, the extrinsic curvature $\theta_t[l_{i_b}] $ is a non-trivial function of all the boundary lengths, determined by solving the bulk Regge equations of motion. (I have used the square-brackets notation to emphasize this non-local character of the dependence.)

Regge theory approximate GR in the following sense.  First, say that a sequence of Regge geometries $g_n$ converges to a Riemannian geometry $g$ if there is map $f$ from one to the other such that for any two points $x$ and $y$, $|d_g(f(x),f(y))-d_{g_n}(x,y)|<\epsilon$, where $d_g(x,y)$ is the distance between $x$ and $y$ in the geometry $g$.  Then, consider a sequence $\Delta_n$ of refinements of a triangulations and a Regge geometry $q_n$ be on the boundary of $\Delta_n$, such that $q_n$ converge to $q$. A discretization is good if 
\be
\lim_{n\to\infty}S_{\Delta_n}[q_n]=S[q]          \label{limregge}
\ee
and I shall assume for the following that the Regge discretization is good.

Let us now come to the quantum theory.  Formally, this can be defined by the truncated transition amplitude
\be
W_{\Delta}[q]=\int \frac{dl_i}{\mu(l_i)}\ e^{\frac{i}{\hbar}I_\Delta[l_i]}           \label{regge}
\ee
where the integral is on the bulk lengths only.  This integral is difficult to define and control for a number of reasons. Among these is the fact that we have little control on the measure and on the triangular inequalities that the lengths $l_i$ must satisfy in order to define a metric, and the fact that the integral is likely to diverge for long or short $l_i$'s. Let me ignore these difficulties for the moment, in order to illustrate the general structure of the theory. In the next Section, I will give a different version of $W_{\Delta}[q]$, where these difficulties are addressed. 

Like the multiple integral \eqref{di}, the truncated transition amplitude \eqref{regge} have two interesting limits. By definition, the $\hbar\to 0$ limit gives the Regge Hamilton function. 
\be
\lim_{\hbar\to 0} (-i\hbar) \log W_{\Delta}[q] = S_\Delta[q].  \label{25}
\ee
If we restore physical units, we must add the Newton constant $1/8\pi G$ in front of the action and \eqref{25} reads
\be
\lim_{l_P\to 0} (-i8\pi l_P^2) \log W_{\Delta}[q_i] = S_\Delta[q_{\partial\Delta}].     
\ee
where $l_P=\sqrt{\hbar G}$ is the Planck length.  Then, taking \eqref{limregge} into account, and in the light of table I, it is reasonable to \emph{define} the quantum gravity transition amplitudes
\be
W[q] = \lim_{n\to\infty} W_{\Delta_n}[q_n]   \label{questo}
\ee
if $\lim_{n\to\infty} q_{n} =q$. (The choice of a particular sequence $\Delta_n$ can be avoided by defining a stricter limit, for $\Delta\to\infty$ in the sense of nets \cite{Rovelli:2010qx}.)  The limit itself,  however, is not of great significance from the perspective of a physicist, since the quantum theory is already defined by its family of approximations \eqref{regge}.  

It is reasonable to expect these approximation to be good in the regime where the corresponding  classical approximations are good, namely where \eqref{limregge} converges fast. This is the regime where the bulk deficit angles are small, namely the regime around flat space.\footnote{The expansion in $n$ should not be confused with the standard perturbative expansion used in QFT. The later is an expansion in the amplitude of the field, keeping all its modes.  The former is an expansion in the number of modes.}

But should we expect the limit \eqref{questo} to actually converge, and converge fast?  A rigorous answer to this equation is missing, but there are circumstantial arguments that indicate that this could be the case in some regimes. The Regge approximation becomes exact on flat space. That is, the Regge action and the continuous Einstein-Hilbert action are equal for a flat geometry.  It follows that if the geometry is flat, the Regge discretization is ``topological" in the sense that it is invariant under a refinement of the triangulation. This very peculiar property was called Ditt-invariance in \cite{Rovelli:2011fk} from Bianca Dittrich, who has emphasized it in her work with Benjamin Bahr \cite{Dittrich:2008pw,Bahr:2011uj,Bahr:2009qc}.  When the boundary data are near flatness, in the sense mentioned, a refinement of the triangulation becomes therefore irrelevant.  More precisely, let $q_n$ be a sequence of boundary Regge metrics converging to a 3-geometry $q$ such that $g_q$ is close to a \emph{flat} geometry. Then one may expect that in this regime $W_{\Delta_n}[q_n]$ converges fast, and therefore a small $n$ is sufficient to give a good approximation to the physical amplitudes. 

\section{Spinfoams}

In covariant loop quantum gravity one defines truncated transition amplitudes which are functions of boundary data \cite{Rovelli:2010wq,Rovelli:2011eq}. These  are defined by the EPRL-FK-KKL dynamics \cite{Engle:2007wy,Freidel:2007py,Kaminski:2009fm}. Explicitly, they are given by
\be
W_{\cal C}(h_l)=\int_{SU(2)} dh_{vf} \ \prod_f \delta(h_f)\ \prod_v A(h_{vf})      \label{ta}
\ee
where the vertex amplitude is given by 
\be
A(h_{ab})=\sum_{j_{ab}}\int_{SL(2C)} \hspace{-.8em} dg_a \ \prod_{ab} {tr}_{\!j_{a\!b}} [h_{ab}Y_\gamma^\dagger g_ag_b^{-1}Y_\gamma ].              \label{ta2}
\ee 
I refer to  \cite{Rovelli:2011eq} for the notation.  These equations define loop quantum gravity. They can be seen as a version of \eqref{regge} that addresses the difficulties associated to \eqref{regge}. In particular, \eqref{ta} is ultraviolet finite and admits a deformed version \cite{Han:2010pz,Fairbairn:2010cp,Han:2011vn} where the amplitude is indeed finite. The measure (from which convergence depends \cite{Crane:2001as}) is fixed by gauge covariance \cite{Bianchi:2010fj} and the triangular inequalities are  implemented by the tensor structure of the $SU(2)$ representations. The theory can be coupled to fermions and Yang-Mills fields \cite{Bianchi:2010bn}.

In \eqref{ta}, ${\cal C}$ is a two-complex.  The transition amplitude depends on $L$ $SU(2)$ elements $h_l$ associated to the $L$ links $l$ of the graph that bounds ${\cal C}$ and is therefore an element $W$ of the boundary Hilbert space $H_{\partial{\cal C}}$  \cite{Rovelli:2011eq}. Semiclassical states $\psi$ in $H_{\partial{\cal C}}$  can be associated to discretized 3d geometries $q_\psi$ formed by glued polyhedra \cite{Freidel:2010aq,Bianchi:2010gc}. In particular, these geometries can be Regge geometries. 

In a remarkable series of works \cite{Barrett:2009mw,Barrett:2009gg,Conrady:2008ea,Magliaro:2011dz}, evidence has piled up that \eqref{ta} converges to the Regge Hamilton function $S_{\Delta}[q]$ in an appropriate classical limit, if the two-complex $\cal C$ is the two-skeleton of the dual of $\Delta$.  We can therefore compose a table similar to the one in Section I.  

\begin{table}[h]
  \centering 
\multirow{2}{10mm}{\begin{sideways}\parbox{20mm}{$\xrightarrow{ \hspace*{2em}{\rm Continuous\ limit} \hspace*{1em}}$}\end{sideways}}
  \begin{tabular}{@{} ccc @{}}
   \parbox{2.9cm}{\scriptsize  Quantum gravity \\  \emph{Transition amplitudes}\\ $W({h_l})$} \hspace*{1em}& $\xrightarrow{j\to\infty}$ &\hspace*{1em} \parbox{2.5cm}{\scriptsize General relativity\\ \emph{Hamilton function} \\ $S[q]$} \\[4mm]
{\begin{sideways}\parbox{10mm}{ $\xrightarrow{{\cal C}\to\infty}$}\end{sideways}} \hspace*{1em}&  &\hspace*{1em} {\begin{sideways}\parbox{10mm}{ $\xrightarrow{{\Delta}\to\infty}$}\end{sideways}}  \\[2mm]
   \parbox{2.5cm}{ \scriptsize LQG amplitudes \\ on a two-complex $W_{\cal C}(h_l)$}\hspace*{1em} & $\xrightarrow{j\to\infty}$ & \hspace*{1em}\parbox{2cm}{\scriptsize Regge theory\\  $S_\Delta(q)$} \\ 
  \end{tabular}\\[.2cm]
\parbox{35mm}{$\xrightarrow{ \hspace*{4em}{\rm Classical\ limit} \hspace*{4em}}$}
  \label{tab:label}
  \caption{Continuous and classical limits in quantum gravity.}
\end{table}

Thus, the LQG transition amplitudes on a two-complex define a family of approximations to the full theory, and the classical limit of each of these (for $\cal C$ the two-skeleton of $\Delta^*$) is given by Regge gravity on $\Delta$. 
An analog table could be written for lattice QCD.

There are some peculiar features of quantum gravity, still, that distinguish it for the case in Table I, as well as from QCD.  

The first is that the three-metric $q$ cannot be fully diagonalized. A maximal set of commuting operators in $H_{\partial{\cal C}}$ is formed by areas and volumes of the polyhedra, but these quantities do not suffice to determine the geometry of the glued polyhedra. The full set of data that determine this geometry is formed by operators that do not commute.  Therefore classical 3 metrics can only be associated to semiclassical states in $H_{\partial{\cal C}}$.  To have semiclassical states --where fluctuations are small with respect to expectation values-- we need large quantum numbers. In particular, we need large spins, and therefore large distances compared to the Planck scale.  Physically, this simply means that at small scale the geometry of space cannot be Riemannian: it is a quantized geometry, and there is a Heisenberg uncertainty preventing the full 3-geometry to be sharp.  In other words, the $\hbar\to 0$ limit of the theory is also necessarily a large distance limit. At short scale, quantum gravity does not have a proper continuous limit taking it to classical GR. This was of course expected on physical grounds.\footnote{There is a formal way to take the classical limit without sending the dimensions of the individual polyhedra to infinity.  Since the eigenvalues of the geometrical quantities are proportional to (powers of) the Immirzi parameter $\gamma$, one can formally take $\gamma$ to zero in order to explore the classical limit at fixed boundary triangulation and at fixed boundary size.  The $\gamma\to 0$ limit has been studied in \cite{Magliaro:2011qm,Magliaro:2011dz,Bianchi:2011fk}.}

A second essential difference between quantum gravity and other theories defined via a discretization such as QCD is the absence of a coupling constant to be tuned to a critical value in order to define the continuous theory. This has been illustrated in detail in \cite{Rovelli:2011fk} and I refer the reader to that paper for a full discussion.

Individual amplitudes (\ref{ta}-\ref{ta2}) can be obtained as Feynman amplitudes of a proper QFT, using the group-field-theory formalism \cite{Geloun:2010vj}.  By separating the terms with vanishing spins and reinterpreting them as defined on sub-two complexes, it may be possible to re-express the limit as a series \cite{Rovelli:2010qx}. This observation, and the analogy with the standard Feynman expansion reinforces the interpretation of the expansion in $n$ as a perturbative expansion.

Finally, the formal argument presented at the end of the last section suggesting fast convergence in $n$ has received some circumstantial support in the context of the amplitude \eqref{ta}: the amplitudes computed  for boundary data sufficiently close to flat space converge rapidly to the correct classical limit already at very low $n$; see \cite{Bianchi:2011fk,Vidotto:2010kw}. The efficacy of the formalism has been proven by the way the old Barrett-Crane model has been ruled out \cite{Alesci:2008ff,Alesci:2007tg,Alesci:2007tx}.

\section{Conclusion} 

Quantum gravity is often confusing.  Field operator insertions in the path integral, which are the main tool for analyzing conventional QFT, are uninteresting in quantum gravity, due to diff-invariance.    The discretization used to define boundary amplitudes is often confused with the quantum discreteness of space. The general structure of a background-independent quantum theory is different from that of a convention QFT.  What are good observables in quantum gravity, and how do we describe evolution? See for example \cite{Giddings:2005id,Tsamis:1989yu,Hartle:1992as,Witten:2001kn,Rovelli:2001bz,Dolby:2004ak,Banks:2002wr,Rovelli:1990ph,Hamber:1993rb,Ashtekar:1993wb,Page:1983uc,Strominger:2001pn,Marolf:1994nz,Wald:1993kj,Kuchar:1991qf} and \cite{Anderson:2010xm} for an overview. 

I have given a tentative overall picture of the structure of the theory, the observables, and the form of the continuous and classical limits.  The truncated boundary transition amplitudes  (\ref{ta}-\ref{ta2})  are the tool for extracting physics from the theory.  From these quantities one can derive standard observables, for instance for analyzing cosmological evolution \cite{Bianchi:2010zs}, or particle scattering \cite{Bianchi:2006uf}, where the old idea that the gravitons live on the non-perturbative quantum states (e.g.\,\cite{Iwasaki:1992qy}) is realized concretely by having the quantum excitations on the nodes of the boundary state. 

These quantities admit two distinct limits. In the classical limit where quantum effects are disregarded, they converge to the Hamilton function of a truncation of GR. In the continuum limit, we expect them to converge to the transition amplitudes of the full theory.  If the present indications are confirmed, there should be a regime in $q$ where the convergence is rapid and therefore the truncation can work as an effective expansion. The expansion parameter is $g_q$'s deviation from flatness.

The truncation introduced by $\cal C$ (or $\Delta$) should not be confused with the physical quantum discreteness of the geometry.  The quantum discreteness of the geometry is the fact that the geometrical \emph{size} of the cells of the complex takes discrete values. It disappears in the semiclassical limit, where the theory is studied at distances large with respect to the Planck scale, while it persists in the continuum limit, with an arbitrary large two-complex.  In other words, no refinement of the cellular complex can make the size of the cells go smoothly to zero, because geometry is physically discrete at the Planck scale.  This is the most characteristic aspect of quantum gravity. 

Finally, not much is known about the effect of the radiative corrections on this structure (for partial results, see \cite{Perini:2008pd,Geloun:2010vj,Krajewski:2010yq,Bonzom:2011br}). These are finite in the deformed version of  \eqref{ta} \cite{Han:2010pz,Fairbairn:2010cp} but this does not make them irrelevant. The main open problem in quantum gravity, I think, is to study their effect on the convergence of the continuous limit.

\vfill

\bibliographystyle{utcaps}
\bibliography{BiblioCarlo}
\end{document}